\documentclass[a4paper,11pt,hyper]{JHEP3}
\usepackage{amsfonts,latexsym,graphicx,epsfig,amssymb,amsmath,mathrsfs}

\newcommand{\bm}[1]{\hbox{\boldmath{$#1$}}}
\newcommand{\sbm}[1]{\hbox{\boldmath{\scriptsize$#1$}}}
\newcommand{\dd}{{\rm d}}
\newcommand{\Mp}{M_{\rm pl}}
\newcommand{\WQ}{W_{\rm QFT}}
\newcommand{\WNL}{\delta W_{\rm NL}}
\newcommand{\WI}{W^{(2)\,-1}}
\newcommand{\cp}{{\rm cyclic~perms}}

\title{Inflation and deformation of conformal field theory}
\author{Jaume Garriga and Yuko Urakawa\\
Departament de F{\'\i}sica Fonamental i Institut de Ci{\`e}ncies del Cosmos, 
Universitat de Barcelona,
Mart{\'\i}\ i Franqu{\`e}s 1, 08028 Barcelona, Spain}

\abstract{It has recently been suggested that a strongly coupled phase of inflation may be described
holographically in terms of a weakly coupled quantum field theory
 (QFT). Here, we explore the possibility that the wave function of an
 inflationary universe may be given by the partition function of a
 boundary QFT. We consider the case when the field theory is a small
 deformation of a conformal field theory (CFT), by the addition of a
 relevant operator $O$, and calculate the primordial spectrum
 predicted in the corresponding holographic inflation
 scenario. Using the Ward-Takahashi identity  
 associated with Weyl rescalings, we derive a simple relation between
 correlators of the curvature perturbation $\zeta$ and 
 correlators of the deformation operator $O$ at the boundary. This is
 done without specifying the bulk theory of gravitation, so that 
 the result would also apply to cases where the bulk dynamics is
 strongly coupled. We comment on the validity of the Suyama-Yamaguchi inequality, 
 relating the bi-spectrum and tri-spectrum of the curvature perturbation.}

\preprint{}

\maketitle

\begin{document}


\section{Introduction}   \label{Sec:Introduction}

The inflationary paradigm is very successful in explaining the initial conditions of our Big Bang universe.
Inflation can be driven by a light scalar field slowly rolling down a potential, and in this case the quantum fluctuations of the scalar field may yield a primordial source of the later structure formation. To verify inflation, and perhaps
uncover a more concrete realization of it, it is important to analyze the predicted primordial fluctuations 
in different models, so they can be compared with observations~\cite{Ade:2013rta}.

In the conventional approach, models of inflation are investigated under the assumption that the fields are weakly coupled, so that their quantum effects can be perturbatively examined. However, since inflation is
expected to take place at a high energy scale, this assumption may not be realized.
The gauge/gravity duality~\cite{Maldacena1997, GKP, Witten1998} states that the dynamics of asymptotically AdS
spacetimes is dual to a CFT at its conformal boundary. AdS/CFT relates weak coupling on one side of the duality to 
strong coupling in the other side. As it stands, such duality cannot be
applied to inflationary cosmology, where the spacetime is similar to de
Sitter rather than AdS. Nonetheless, by analogy, there have been
several suggestions that a $(d+1)$-dimensional inflationary
evolution may be dual to a quantum field theory (QFT) on a
$d$-dimensional space {of Euclidean signature}. Following the work by
Strominger~\cite{Strominger, Strominger2} and Witten~\cite{Witten}, the
possible duality between gravity in de Sitter space and a CFT has been
investigated in Refs.~\cite{Bousso:2001mw, Harlow:2011ke, Anninos:2011ui}. A
holographic description of quasi-de Sitter spacetime would provide a new
approach for analyzing the origin of cosmological
perturbations~\cite{Strominger2, Maldacena02, Seery:2006tq}, allowing us to consider a
strongly coupled phase of inflation. This would correspond to the weakly
coupled phase of the boundary theory, which can be treated
perturbatively.

McFadden and Skenderis \cite{MS_HC09, MS_HC10, MS_HCob10, MS_NG,
MS_NGGW, BMS} have recently focused on the
so-called domain wall/cosmology correspondence. This is based on
the observation that domain wall solutions can be mapped into
cosmological solutions by analytic continuation \cite{Skenderis:2006jq,
Skenderis:2006fb}. First they describe the gravitational field of the
domain wall solution in terms of the dual QFT on the boundary based on the
gauge/gravity duality, using the holographic
renormalization group method~\cite{Verlinde, Kostas_RG, Kostas_RG02}. 
Then, they perform the analytic continuation to
transform the domain wall solution into the cosmological spacetime
solution. Skenderis and McFadden show that the mapping can be
extended to include cosmological perturbations. 
The results presented in Refs.~\cite{MS_HC09, MS_HC10, MS_HCob10,
MS_NG, MS_NGGW, BMS} are rather robust and amenable to comparison with upcoming cosmological
observations~\cite{MS_HCob10, EFMS, Dias}.

Motivated by these developments, we further investigate the holographic
description of the inflationary spacetime.  Here, we take as our starting point the assumption
that the wave function of the $(d+1)$-dimensional bulk
gravitational field of the inflationary spacetime is given by the generating functional of the
dual QFT on the $d$-dimensional boundary~\cite{Strominger2, Witten,Maldacena02,vdS}. (See also Refs.~\cite{
Shiu, Mata:2012bx}.)  The bulk dynamics is then supposed to be dual
to a QFT on the Euclidean boundary and the RG flow describes the
bulk evolution. Under this assumption, we will formally
derive the relation between the primordial fluctuation in the bulk
inflationary spacetime and the dual QFT on the boundary. To introduce a
deviation from the exact de Sitter symmetry in the bulk, we will
consider a deformation of the CFT on the boundary. 
Our approach is manifestly independent of the bulk dynamics. Rather, the theory is
defined in terms of the dual QFT.

The outline of this paper is as follows. In Sec.~\ref{Sec:Setup}, we describe our
setup. In
Sec.~\ref{Sec:GF}, using the Ward-Takahashi identity associated with
the Weyl transformation, we derive the relation between the wave
function and the boundary QFT operators. In Sec.~\ref{Sec:PS}, using
the formulas derived in the preceding sections, we give the formal
expression of the primordial spectra described by the correlation
functions of the boundary QFT operators. We also investigate the
validity of the Suyama-Yamaguchi inequality, which is known to hold for
a wide class of weakly coupled inflation models.

\section{Preliminaries}  \label{Sec:Setup}
The cosmological spacetime metric can be given in ADM formalism as
\begin{align}
 & \dd s^2 =  - N^2 \dd t^2 + h_{ij} \left( \dd x^i + N^i \dd t
 \right) \left( \dd x^j + N^j \dd t \right)\,. \label{Exp:metricb}
\end{align}
Here we shall restrict attention to the situation where the metric is asymptotically de Sitter.
In the semiclassical picture, this would correspond to the case a period of slow roll inflation is followed by 
$\Lambda$ domination. In this case, the future conformal boundary is space-like and will be thought of 
as home to a QFT.

Our starting point is the assumption that the wave function of
the bulk gravitational field is related to the generating functional of
the boundary QFT
\begin{align}
 & \psi_{\rm bulk}[h,\phi] \propto Z_{\rm QFT}[h,\phi]\,, \label{duality}
 \end{align}
where the generating functional $Z_{\rm QFT}$ is given by
\begin{equation}
Z_{\rm QFT} [h,\phi]  =  e^{-W_{\rm QFT}[h,\phi]} =  \int D \chi\,
\exp \left(  - S_{\rm QFT} [\chi,\, h,\, \phi]\right)   \, \label{Exp:Z}.
\end{equation}
Here $\psi_{\rm bulk}$ denotes the wave function of the bulk and $\chi$ denotes boundary
fields, for which the metric $h_{ij}$ and the inflaton $\phi$ act as sources (the indices in $h_{ij}$ will be omitted when unnecessary). 
Note that $h$ and $\phi$ correspond to the values of the metric and
inflaton at the future boundary, where any finite co-moving separation
corresponds to an infinite physical wavelength in the bulk
theory.

Since the wave function $\psi_{\rm bulk}$ is generally complex,
$W_{\rm QFT}$ will have real and imaginary part, which means that this  
is not the effective action of a standard unitary Euclidean field theory. In the regime where
gravity is semiclassical, we have 
\begin{equation}
\psi_{\rm bulk} \approx |\psi_{\rm bulk}| e^{iS_{\rm bulk}},
\label{bulk}
\end{equation}
where $S_{\rm bulk}$ is the classical bulk action as a function of the boundary fields.
For long wavelength modes, the modulus is approximately constant in time, while the classical action $S_{\rm bulk}$ is extensive in the 
space-time volume and grows unbounded with the scale factor $a$.
In the dual picture, the scale factor will play the role of the renormalization scale $\mu$, so the divergent terms in
$\WQ$ are in fact purely imaginary. Because of that, some imaginary counterterms will be necessary in $S_{\rm QFT}$, and
one may prefer to define the generating functional as \cite{JG09}
\begin{equation}
Z_{\rm QFT} = e^{i\bar W_{\rm QFT}} = \int D \chi\,
\exp \left(  i \bar S_{\rm QFT} [\chi,\, h,\, \phi]\right), \label{notation2} 
\end{equation}
where 
\begin{align}
 & \bar W_{\rm QFT}= i W_{\rm QFT},\  \quad \quad \bar S_{\rm QFT} = i S_{\rm QFT}. \label{Exp:Ss}
\end{align}
In this notation, the counterterms in $\bar S_{\rm QFT}$ would be purely real. Here, we will simply follow the notation 
in Ref.~\cite{Harlow},  where Eq.(\ref{Exp:Z}) is used with the understanding that $S_{\rm QFT}$ is necessarily complex. The interchange between the two notations is of course trivial.

\subsection{Deformed CFT}

We shall concentrate on the case where the QFT is a small deformation of
a CFT with
\begin{align}
 & S_u [\chi] = S_{\rm CFT} [\chi] + \int \dd^d \bm{x} \sqrt{h} \, u  \,O(\bm{x}) \,. \label{Exp:Su}
\end{align} 
The first term preservers the conformal symmetry, and 
$O(\bm{x})$ is a QFT operator with the scaling dimension
$\Delta$. Introducing 
\begin{align}
 & \lambda \equiv \Delta - d\,,
\end{align}
the case  $\lambda=0$ corresponds to a marginal deformation, which changes the theory into another CFT. An operator
with $\lambda>0$ corresponds to an irrelevant deformation and an operator with $\lambda<0$ corresponds to a 
relevant deformation~\cite{Hollowood:2009eh}. An operator with 
$\lambda \neq 0$ introduces a renormalization group flow which is
characterized by the beta
function. This is defined for a dimensionless coupling constant 
\begin{equation}
g(\mu) \equiv \mu^{\lambda_c} u(\mu)
\end{equation}
as
\begin{align}
 & \beta \equiv  \frac{\dd g(\mu)}{\dd \ln \mu}\,,
\end{align}
where $\mu$ is the energy scale and we have introduced 
\begin{align}
 & \lambda_c \equiv \delta - d\,.
\end{align}
Here, $\delta$ is the classical scaling dimension, which is in general
different from $\Delta$. Noticing the fact that the coupling always appears in the combination
$\mu^{-\lambda_c} g(\mu)$ in the classical action, we can express the beta
function as
\begin{align}
 & \beta = \lambda_c g  + \beta^{\rm quant}\,,
\end{align}
where integrating the quantum corrections gives rise to the deviation
from the classical scaling. We assume that
the theory approaches a fixed point at high energies, so that $u$
approaches 0. 
Conformal symmetry is then recovered. 
The coupling constant $g$ of the boundary theory is interpreted as the boundary value of the inflaton $\phi$, 
\begin{equation} 
g=\phi/\Mp \label{inflaton}
\end{equation}
where we divided the inflaton by the Planck scale $\Mp$ to keep $g$
dimensionless. 

If the energy scale $\mu$ is identified with the 
cosmological scale factor $\mu \propto a$, we can rewrite the beta function
as $\beta \simeq (\dd /\dd \ln a)  \phi/\Mp$. Once we assume
the field equation for the bulk gravitational field, we can express the
deformation parameters $\lambda$ and $\beta$ in terms of the so-called
slow-roll parameters, which describe the deviation from the exact de
Sitter space. For instance, using the Friedmann equation, we can express
$\beta$ in terms of the slow-roll parameter 
$\varepsilon\equiv H (\dd /\dd \ln a) 1/H $ as 
$\beta^2 \propto \varepsilon$~\cite{vdS, Shiu}. However, in the strongly
coupled limit of bulk gravity, such relations need not apply. 
In this paper, we will not make any reference to bulk equations of motion. 
Our arguments are based on the assumption that the boundary theory is dual to
the (perhaps strongly coupled) bulk theory through Eq. (\ref{duality}). This provides a direct 
relation between the long wavelength correlators  of the bulk gravitational field and the correlators of the field theory 
operator $O$. The parameters $\beta$ and $\lambda$ will be fixed by
solving the RG flow from the UV fixed point to the IR fixed
point~\cite{BMS, Klebanov:2011gs}. Since our aim is to give a formal expression of the
primordial spectrum in terms of boundary QFT operators, we simply
assume that $\beta$ and $\lambda$ are given functions of the energy scale.

\subsection{The wave function}

Throughout this paper, we shall restrict attention to the scalar component of the boundary fields $h_{ij}$ and $\phi$.
In a bulk description, we adopt the gauge condition $\delta \phi=0$ and 
\begin{align}
 & h_{ij} = a^2(t) e^{2\zeta} \delta_{ij}\,,  \label{Exp:gauge}
\end{align}
where we neglected the tensor modes. We normalize the scale
factor $a(t)$ as $a(t_c)=1$ where $t_c$ is the coordinate $t$ which
corresponds to the UV cutoff scale of QFT. In the following, we
calculate the primordial fluctuation at a fixed probe brane with
$t=t_c$. For notational simplicity, we abbreviate the coordinate $t$ in
the arguments of fields.    

From  (\ref{Exp:Z}), we have
\begin{align}
 & \psi_{\rm bulk}[\zeta] = A\, Z_{\rm QFT}[\zeta] =A\, 
e^{- \WQ[\zeta]}\,, \label{Exp:psi}
\end{align}
where we wrote a normalization constant $A$ explicitly.
We may now calculate $n$-point functions of
the gravitational field perturbation in terms of the QFT correlators. 
From the wave function $\psi_{\rm bulk}[\zeta]$, we obtain the probability density
\begin{align}
 P[\zeta] = \left| \psi_{\rm bulk}[\zeta] \right|^2 
 = |A|^2\, e^{-(\WQ[\zeta] + \WQ^*[\zeta]) }\,.\label{hdist}
\end{align}
Once we obtain the probability density function $P[\zeta]$, we can calculate the $n$-point
functions for $\zeta$ on the boundary as
\begin{align}
 & \langle \zeta(\bm{x}_1) \zeta(\bm{x}_2)  \cdots \zeta(\bm{x}_n) \rangle = 
 \int D \zeta\, P[\zeta]\, \zeta(\bm{x}_1) \zeta(\bm{x}_2)  \cdots \zeta(\bm{x}_n)\label{cor}
\end{align}
where $D \zeta$ is the integration measure. In weakly coupled Einstein
gravity, the bulk distribution is given by a nearly Gaussian distribution for the
variable $\zeta$, accompanied by a measure which is linear in $\zeta$. 
On the other hand, our aim here is to proceed without reference
to the explicit  form of the bulk theory. Hence, after postulating the
correspondence (\ref{Exp:psi}) for the wave function, we still need some
justification for choosing $D \zeta$ instead of, say, $D f(\zeta)$,
where $f$ is an arbitrary function. As we shall see, the holographic
distribution (\ref{hdist}) becomes independent of $\zeta$ as we approach
the conformal fixed point. This limit corresponds to de Sitter space,
where the variable $\zeta$ is pure gauge. We therefore expect that all
values of $\zeta$ should become equally probable as the conformal fixed
point is approached. This will only be realized if we choose $f$ to be
linear in $\zeta$. In this case, the measure is invariant under the Weyl
rescalings, since as we will see in the next section, the Weyl
rescalings simply shift $\zeta$ as $\zeta \to \zeta -\alpha$, where
$\alpha$ is an arbitrary function of position.

We determine
the normalization constant $A$, adopting the normalization condition:
\begin{align}
 & \int D \zeta P[\zeta] =1\,. \label{Cond:norm}
\end{align}
Eliminating the background contribution $\WQ[\zeta=0]$ by the
redefinition of $A$, $P[\zeta]$ is given by 
\begin{align}
 & P[\zeta]= |A|^2 e^{- \zeta \cdot W^{(1)} - \frac{1}{2} \zeta \cdot W^{(2)} \cdot
 \zeta - \WNL[\zeta] }\,,
\end{align}
where we defined the vertex function $W^{(n)}(\bm{x}_1,\, \cdots,\,\bm{x}_n)$ for $n\geq 1$ as
\begin{align}
 & W^{(n)}(\bm{x}_1,\, \cdots,\,\bm{x}_n)\equiv  2\, {\rm Re} \left[ \frac{\delta^n \WQ[\zeta]}{\delta\zeta(\bm{x}_1)
 \cdots \delta \zeta(\bm{x}_n)} \right]\,. \label{Def:Wn}
\end{align}
For a later use, we discriminated the non-linear interactions,
introducing $\WNL[\zeta]$ which is defined as
\begin{align}
 & \WNL[\zeta] \equiv  \sum_{n=3} \frac{1}{n!} \int \dd^d \bm{x}_1 \cdots \int \dd^d \bm{x}_n\,
 W^{(n)}(\bm{x}_1,\,\cdots\,,\bm{x}_n) \zeta(\bm{x}_1) \cdots
 \zeta(\bm{x}_n) \,.  \label{Exp:WNL}
\end{align}
Here and hereafter we use an abbreviated notation:
\begin{align}
 & \phi \cdot \psi \equiv \int \dd^d \bm{x} \phi(\bm{x}) \psi(\bm{x})\,, \\
 & \phi \cdot M \cdot \psi \equiv \int \dd^d \bm{x}_1 \!\int\! \dd^d \bm{x}_2\, \phi(\bm{x}_1)
 M(\bm{x}_1,\, \bm{x}_2) \psi(\bm{x}_2)\,
\end{align}
for functions $\phi(\bm{x})$ and $\psi(\bm{x})$ and a matrix function 
$M(\bm{x}_1,\,\bm{x}_2)$.

First, we consider the case with $W^{(1)}(\bm{x})=0$.
In deriving the $n$-point functions for $\zeta$, it is convenient to
perform the Legendre transform as
\begin{align}
 & \Phi[J] \equiv \int D \zeta\, P[\zeta] e^{- \zeta \cdot J}
 = |A|^2 \int D \zeta\, e^{- \frac{1}{2}\zeta \cdot
 W^{(2)} \cdot \zeta  - \WNL[\zeta] - \zeta \cdot  J } \,.
\end{align}
We can easily understand that $n$-th derivatives of $\Phi[J]$ in
terms of $J(\bm{x})$ yield $n$-point functions of $\zeta(\bm{x})$. Using the
generating functional for the free field given by
\begin{align}
 & \Phi_0[J] =  |A|^2 \int D \zeta\, e^{- \frac{1}{2}\zeta \cdot
 W^{(2)} \cdot \zeta  - \zeta \cdot J}= |A|^2\, e^{\frac{1}{2} J \cdot \WI \cdot
 J}\,,
\end{align}
where $\WI (\bm{x}_1,\, \bm{x}_2)$ is the inverse matrix of 
$W^{(2)}(\bm{x}_1,\, \bm{x}_2)$, which satisfies 
\begin{align}
 & \int \dd^d \bm{x}\, W^{(2)}(\bm{x}_1,\, \bm{x}) \WI (\bm{x},\,\bm{x}_2) =
 \delta(\bm{x}_1- \bm{x}_2)\,, \label{Def:D}
\end{align}
the generating functional $\Phi[J]$ is recast into
\begin{align}
  \Phi[J] &= e^{-\WNL[\delta/\delta J]}\, \Phi_0[J] = |A|^2
 e^{-\WNL[\delta/\delta J]}\, e^{\frac{1}{2} J \cdot \WI \cdot
 J} \cr &= |A|^2 e^{\frac{1}{2} \delta/ \delta \zeta \cdot
  \WI \cdot \delta/\delta \zeta} e^{-\WNL[\zeta] - \zeta\cdot J} \Big|_{\zeta=0}\,.
\end{align}
On the last equality, we used the following identity:
\begin{align}
 & F[\delta /\delta \zeta] G[\zeta] e^{-\zeta\cdot J} \big|_{\zeta=0} =
  G[\delta/ \delta (-J)] F[-J]\,,
\end{align}
which is valid for arbitrary functionals $F$ and $G$. 
Substituting $|A|$ which is determined by the normalization condition
(\ref{Cond:norm}), the generating functional $\Phi[J]$ is finally given by 
\begin{align}
 & \Phi[J] = \frac{ e^{\frac{1}{2} \delta/ \delta \zeta \cdot
 \WI \cdot \delta/\delta \zeta} e^{-\WNL[\zeta] - \zeta\cdot
  J} \big|_{\zeta=0}}{ e^{\frac{1}{2} \delta/ \delta \zeta \cdot
 \WI \cdot \delta/\delta \zeta} e^{-\WNL[\zeta]} \big|_{\zeta=0}}\,. \label{Exp:PhiJ}
\end{align}
This formula can be understood as follows. We first expand the
exponential terms 
$e^{-\WNL[\zeta] - \zeta\cdot J}$ up to a desired order in
perturbation. A coefficient of a term with $J(\bm{x}_1) \cdots J(\bm{x}_n)$
gives the $n$-point function for $\zeta$. These
correlators can be evaluated after we multiply the derivative operator
$e^{\frac{1}{2} \delta/ \delta \zeta \cdot  \WI \cdot \delta/\delta \zeta}$,
which replaces two $\zeta$s with $\WI$. Since we set $\zeta=0$
after this replacement, only terms with an even number of $\zeta$s can
survive. We can derive the Feynman rules as in the path integral of a
usual quantum field theory. The only difference is that, in the present case,
the vertices  $W^{(n)}$ are generically non-local, depending on $n$ separate points $\bm{x}_i$ which 
are integrated over. Different ways of taking derivatives by
$\delta/\delta \zeta$ correspond to different diagrams. In the diagrammatic description, 
the non-locality of the vertices can be illustrated
by splitting the interaction point into $n$ separated ones, as in
Figs.~\ref{Fg:V3} and \ref{Fg:V4}.  The generating functional $\Phi[J]$ for
the case with $W^{(1)} \neq 0$ can be obtained simply by replacing
$\WNL[\zeta]$ with $\WNL[\zeta]+ W^{(1)}\cdot \zeta$ in the exponents of
Eq.~(\ref{Exp:PhiJ}). Following the conventional quantum field
theory, to pick up only connected diagrams, we define the $n$-point
functions as
\begin{align}
 & \langle \zeta(\bm{x}_1) \cdots \zeta(\bm{x}_n) \rangle_{\rm conn}
 \equiv (-1)^n \frac{\delta^n}{\delta J(\bm{x}_1)\cdots \delta
 J(\bm{x}_n)} \ln \Phi[J] \bigg|_{J=0} \, \,,  \label{PsiJ}
\end{align}
where we put the suffix ``conn'' to emphasize including only connected
diagrams.

\section{Derivatives of the generating functional}  \label{Sec:GF}
In this section, using the Ward-Takahashi identity associated with the
Weyl transformation, we express the vertex function 
$W^{(n)}(\bm{x}_1,\,\cdots,\, \bm{x}_n)$ in terms of field theory correlators. The classical
scalings of the metric $g_{ij}$ and the operator $O$ are given by
\begin{align}
 & g_{ij}(\bm{x}) \to e^{-2\alpha(\sbm{x})} g_{ij}(\bm{x})\,, \qquad 
 O(\bm{x}) \to  e^{\Delta \alpha(\sbm{x})}
 O(\bm{x})\equiv O_{\alpha}(\bm{x})\,.  \label{Exp:Weyl}
\end{align}
In the gauge with Eq.~(\ref{Exp:gauge}), the Weyl transformation of the metric is expressed as
\begin{align}
 & \zeta(\bm{x}) \to \zeta(\bm{x}) -\alpha(\bm{x}) \equiv \zeta_\alpha(\bm{x})\,.
\end{align}
In the following, we will show that the generating functional
$\WQ[\zeta]$ and also the vertex function $W^{(n)}(\bm{x}_1,\, \cdots,\, \bm{x}_n)$ are related to correlation functions of $O$ by a simple
relation. Since we fixed the slicing of the $t$-constant surface requesting
$\delta \phi=0$, the coupling constant $u$ is homogeneous and will not
be affected by the Weyl transformation. 

\subsection{The Ward-Takahashi identity}
Under the Weyl transformation (\ref{Exp:Weyl}), the action $S_u$
transforms as
\begin{align}
 S_u[\zeta,\, \chi] \to 
 S_u[\zeta_\alpha,\,\chi_\alpha] &= S_{\rm CFT} [\zeta_\alpha,\,\chi_\alpha] + \int \dd^d
  \bm{x} e^{d \zeta_\alpha (\sbm{x})}  u\, O_{\alpha} (\bm{x})  \cr
  & =   S_{\rm CFT} [\zeta,\, \chi] + \int \dd^d \bm{x} e^{d \zeta(\sbm{x})}  e^{(\Delta - d)
 \alpha(\sbm{x})} u\, O(\bm{x}) \cr
  &  = S_u[\zeta,\, \chi] + \delta S_\alpha [\zeta,\, \chi]  \,,  \label{Exp:trans}
\end{align}
where we defined
\begin{align}
 & \delta S_\alpha [\zeta,\, \chi]\equiv  \int \dd^d \bm{x} e^{d
 \zeta(\sbm{x})}  
 \left[  e^{\lambda \alpha(\sbm{x})} -1 \right] u\, O(\bm{x})\,.
\end{align}
On the second equality, we noted that $S_{\rm CFT}$ is invariant
under the Weyl transformation. (The conformal
invariance only implies the invariance under the Weyl transformation with the
specific $\bm{x}$-dependent parameter $\alpha(\bm{x})$. To be more
precise, $ S_{\rm CFT}$ is assumed to preserve the Weyl invariance
rather than the conformal symmetry.) Noticing the fact that the
integration of the generating functional is independent of the choice of
integration variable, we express $Z_{\rm QFT}$ as
\begin{align}
 & Z_{\rm QFT} =  \int D \chi\, e^{- S_u[ \zeta_\alpha,\,\chi]}=\int D \chi_\alpha\, e^{- S_u[\zeta_\alpha,\,
 \chi_\alpha]}\,.  
\end{align}
Using Eq.~(\ref{Exp:trans}), we can rewrite this relation as
\begin{align}
 &   \int D \chi\, e^{- S_u[ \zeta - \alpha,\,\chi]}=\int D \chi\, {\cal J}[\alpha] e^{-  \{S_u[\zeta,\, \chi] + \delta S_\alpha
 [\zeta,\, \chi] \}}\,, \label{WT}
\end{align}
where ${\cal J}[\alpha]$ denotes the Jacobian of the integration measure, 
{\it i.e.,} $D \chi_\alpha = {\cal J}[\alpha] D\chi$. Equation (\ref{WT})
provides the so-called Ward-Takahashi identity associated with the Weyl
transformation.

The Jacobian ${\cal J}[\alpha]$
can be obtained by considering the transformation of the boundary fields
$\chi$ under the Weyl transformation. A change of the integration
measure $D\chi$ is known to yield the Weyl anomaly (see for instance
Ref.~\cite{Fujikawa:1980rc}). In odd dimensions, we can employ
${\cal J}[\alpha]=1$, since the Weyl anomaly is absent. In even
dimensions, since the anomaly is present, ${\cal J}[\alpha]$ can be
different from 1 (see for instance Ref.~\cite{BD} and references therein).
Therefore, here, we focus on odd
dimensions with ${\cal J}[\alpha]=1$, leaving a study of the case with
${\cal J}[\alpha] \neq 1$ for future issue.

Next, we operate
$\delta^n/\delta \alpha(\bm{x}_1) \cdots \delta \alpha(\bm{x}_n)$ on the
both sides of Eq.~(\ref{WT}) and set all $\alpha(\bm{x})$s to $0$.
Noticing the fact that the path integral of the left-hand side of
Eq.~(\ref{WT}) includes $\alpha(\bm{x})$ only in the form of 
$\zeta(\bm{x}) - \alpha(\bm{x})$, we can obtain the Ward-Takahashi identity as
\begin{align}
 &  (-1)^n \frac{\delta^n}{\delta \zeta(\bm{x}_1) \cdots \delta
 \zeta(\bm{x}_n)} \int D \chi\, e^{- S_u[ \zeta,\,\chi]} \cr
 & \qquad \quad  =\frac{\delta^n}{\delta \alpha(\bm{x}_1) \cdots \delta
 \alpha(\bm{x}_n)} \int D \chi\, e^{-  \{S_u[\zeta,\, \chi] + \delta S_\alpha
 [\zeta,\, \chi] \}}  \bigg|_{\alpha=0}\,, \label{WTodd1}
\end{align}
where we noted that since the integration measure does
not change under the Weyl transformation induced by a change of
$\zeta$, $D\chi$ commutes with $\delta/\delta \zeta$. The left hand
side of Eq.~(\ref{WTodd1}) is nothing but the differentiation of the
generating functional $Z_{\rm QFT}[\zeta]$ in terms of
$\zeta(\bm{x})$. Dividing Eq.~(\ref{WTodd1}) by $Z_{\rm QFT}[\zeta]$ and
sending $\zeta$ to 0, we obtain
\begin{align}
 & \frac{1}{Z_{\rm QFT}[\zeta]} \frac{\delta^n}{\delta \zeta(\bm{x}_1) \cdots \delta
 \zeta(\bm{x}_n)} Z_{\rm QFT}[\zeta] \bigg|_{\zeta=0} 
= (-1)^n \left\langle \frac{\delta^n}{\delta \alpha(\bm{x}_1) \cdots \delta
 \alpha(\bm{x}_n)}  e^{-   \delta S_\alpha [\zeta,\, \chi]}
 \bigg|_{\alpha=0} \right\rangle_u  \label{WTodd}
\end{align}
where we defined
\begin{align}
 & \langle X[\chi] \rangle_u \equiv \frac{1}{Z_{\rm QFT}[\zeta=0]} \int D \chi\, e^{-
 S_u[\zeta=0,\, \chi]} X[\chi] \,.
\end{align}
To emphasize $u$-dependence, we put the index $u$.

After a straightforward calculation, we can express the right hand side of
Eq.~(\ref{WTodd}) in terms of the correlation functions of the QFT
operator $O(\bm{x})$ as
\begin{align}
 & \frac{1}{Z_{\rm QFT}[\zeta]}  \frac{\delta^n}{ \delta \zeta(\bm{x}_1)
  \cdots \delta \zeta(\bm{x}_n)} 
 Z_{\rm QFT} [\zeta] \bigg|_{\zeta=0} \cr
 & \quad  = \lambda^n \biggl[ u^n \langle O(\bm{x}_1) \cdots
 O(\bm{x}_n) \rangle_u  \cr
 & \qquad \qquad   \quad - u^{n-1} \left\{ \delta
 (\bm{x}_1 - \bm{x}_2)  \langle O(\bm{x}_2) \cdots O(\bm{x}_n) \rangle_u  + \left( \frac{n(n-1)-2}{2}\,\, \cp \right)  \right\} \cr
 & \qquad \qquad   \quad  + \cdots + (-1)^{n-1} u \, \delta
 (\bm{x}_1 - \bm{x}_2) \cdots \delta(\bm{x}_{n-1} - \bm{x}_n)
 \langle O(\bm{x}_1) \rangle_u \biggr]\,. \label{Exp:Zn}
\end{align}
Using Eq.~(\ref{Def:Wn}) and dropping the terms which do not contribute
to the connected diagram, we obtain the vertex function
$W^{(n)}(\bm{x}_1,\, \cdots,\, \bm{x}_n)$ as
\begin{align}
 W^{(n)}(\bm{x}_1,\, \cdots,\, \bm{x}_n)
 &= - 2 {\rm Re} \left[  \frac{1}{Z_{\rm QFT}[\zeta]} \frac{\delta^n
 Z_{\rm QFT}[\zeta]}{\delta \zeta(\bm{x}_1) \cdots \delta 
 \zeta(\bm{x}_n) } \bigg|_{\zeta=0} \right] \,.  \label{Exp:WQF}
\end{align}
In this way, using the Ward-Takahashi identity, the vertex function 
$W^{(n)}(\bm{x}_1, \cdots, \bm{x}_n)$ can be expressed in terms of the
correlation functions for the QFT operator $O$ and the deformation parameters
$\lambda$ and $u$, which characterize the deviation from the conformal
fixed point. Note that at the fixed point, where $u$ vanishes, the
right-hand side of Eq.~(\ref{Exp:WQF}) totally vanishes.

\subsection{Other approaches}

In the previous subsection, we have related the vertex function for $\zeta$ to the correlation functions
for an operator that lives at the future boundary. This is in contrast
with the approach by McFadden and Skenderis, who considered the
primordial spectra of $\zeta$ in the context of the domain-wall/cosmology correspondence~\cite{MS_HC09, MS_HC10, MS_HCob10, MS_NG,
MS_NGGW, BMS}. The present formalism can be applied also to the domain wall
space, and the distribution function of the bulk field in the
domain-wall space is still given by the same expressions. (For
instance, the formula for the vertex function given by
Eqs.~(\ref{Def:Wn}) and (\ref{Exp:Zn}) still holds.)

Let us now comment on the difference between our approach and the one in
Refs.~\cite{MS_HC09, MS_HC10, MS_HCob10, MS_NG, MS_NGGW, BMS}. 
Since $\phi$ and $h_{ij}$ play the role of the external
fields for $O$ and $T_{ij}$, respectively, we can naively express the wave function for the bulk as
\begin{align}
 & \psi_{\rm bulk}[ \delta \phi,\, \delta h^{ij}]   \cr
 &= \exp \biggl[ - \frac{1}{2 \Mp^2}
 \!\int\! \dd^d \bm{x}_1\! \int\! \dd^d \bm{x}_2 \sqrt{h(\bm{x}_1)} \sqrt{h(\bm{x}_2)} 
 \delta \phi (\bm{x}_1)  \delta \phi(\bm{x}_2) \langle O(\bm{x}_1) O(\bm{x}_2)
 \rangle \cr
 & \quad  - \frac{1}{2} \!\int\! \dd^d \bm{x}_1 \! \int \! \dd^d
 \bm{x}_2  \sqrt{h(\bm{x}_1)} \sqrt{h(\bm{x}_2)}
\delta h^{ij} (\bm{x}_1)  \delta h^{kl}(\bm{x}_2) \langle T_{ij}(\bm{x}_1) T_{kl}(\bm{x}_2)
 \rangle + \cdots  \biggr]\!, \label{Exp:HF}
\end{align}
where $\delta \phi$ and $\delta h^{ij}$ denote the
fluctuations of the inflaton and the gravitational field,
respectively. For comparison, let us comment on how the wave function
$\psi_{\rm bulk}$ can be related to the QFT operator
$O$, by starting with the formula
(\ref{Exp:HF}). Since the longitudinal mode of the gravitational field
and the fluctuation of the inflaton $\delta \phi$ constitute a single gauge
invariant variable, (the longitudinal mode of) the bulk degree of
freedom can be described either by
$\delta \phi$ or by the longitudinal part of the gravitational field on
the fixed gauges. Here, we fixed $\delta \phi=0$ using the gauge degree
of freedom. In the absence of the tensor modes, only the trace part of the energy momentum tensor
contributes in the second term of Eq.~(\ref{Exp:HF}). As is shown in
Ref.~\cite{BMS}, the
Ward-Takahashi identity associated with the Weyl transformation gives the relation between the correlation
functions of the trace of the energy momentum tensor $T$ and the correlation
functions of the QFT operator $O$. In this way, we
can express the wave function for the bulk gravity $\psi_{\rm bulk}$ and
the vertex functions 
$W^{(n)}(\bm{x}_1,\, \cdots,\, \bm{x}_n)$ in terms of the correlation
functions of $O$. On the other hand,  the expression of
Eq.~(\ref{Exp:HF}) is not very well adapted for computing
the correlation functions of the curvature perturbation $\zeta$ because
of the presence of local terms. These are the terms in 
$W^{(n)}(\bm{x}_1,\, \cdots,\, \bm{x}_n)$ which have
at least one delta function $\delta(\bm{x}_i - \bm{x}_j)$ where 
$i, j=1, \cdots, n$. Local terms appear due to the
following two reasons. First, since the energy momentum tensor
depends on the gravitational field, the metric fluctuation appears in the form,
$$
\delta h^{ij}(\bm{x}_1) \delta h^{i'j'}(\bm{x}_1') \delta h^{kl}(\bm{x}_2) \cdots
\left\langle \frac{\delta 
T_{ij}(\bm{x}_1)}{\delta h^{i'j'}(\bm{x}_1')} T_{kl}(\bm{x}_2) \cdots \right\rangle\,, 
$$
which will give the delta function $\delta(\bm{x}_1-
\bm{x}_1')$~\cite{MS_NG, BMS}.  Second, more than one $\zeta$s
can appear in the integration of $\bm{x}$ through the invariant volume
$e^{d \zeta(\sbm{x})} \dd^d \bm{x}$ and 
$\delta h^{ij}= (e^{-2 \zeta(\sbm{x})}-1)\delta^{ij}$. Because of these
local terms, deriving the wave function by inserting the Ward-Takahashi identity
which relates the energy momentum tensor and $O$ in
Eq.~(\ref{Exp:HF}) leads to unnecessarily complicated expressions.

In Refs.~\cite{MS_HC09, MS_HC10, MS_NG, MS_NGGW, BMS}, the authors used the Fefferman-Graham gauge.  In this case,
we would also need to take into account the first term of Eq.~(\ref{Exp:HF}) in our derivations, and then we 
should express the generating functional in terms of the gauge invariant variable $\zeta$. The
resulting correlation functions for $\zeta$ should not depend on the choice
of gauge used in its derivation.

In this paper, we provided a direct way to
express the wave function $\psi_{\rm bulk}$ in terms of the correlation functions
of $O$ without assuming a field equation in bulk. We showed that as is given in Eq.~(\ref{Exp:Zn}), the Ward-Takahashi identity can be
expressed in a way that the wave function 
$\psi_{\rm bulk} \propto Z_{\rm QFT}$ is directly related to the
correlation functions of $O$, which enables us to skip the redundant
process included in the above-mentioned procedure.

\section{The primordial spectra}   \label{Sec:PS}
Once we have obtained the vertex functions,
$W^{(n)}(\bm{x}_1,\, \cdots,\, \bm{x}_n)$, using  Eq.~(\ref{PsiJ}),
which gives the Feynman rule, we can calculate the $n$-point functions
of $\zeta(\bm{x})$. In this section, we derive a general expression of the $n$-point function
for the curvature perturbation $\zeta(\bm{x})$,
whose dual field theory is given by a deformed conformal field
theory. Here, setting $d$ to 3, we consider the case with 
\begin{equation}
 W^{(1)}(\bm{x})= - 2\, {\rm Re} \left[  \lambda u \langle O (\bm{x})
 \rangle_u \right] = 0. \label{that}
\end{equation}
Equation (\ref{that}) is valid at the lowest order in the deformation parameter.
At higher orders, $W^{(1)}$ can take a non-vanishing value. Namely, when the global translation invariance is
 preserved, $W^{(1)}$ should take a constant value. Then, the
 contribution of $W^{(1)}$ to the probability density $P[\zeta]$, given by
$$
\int \dd^3 \bm{x} W^{(1)}(\bm{x}) \zeta(\bm{x}) =  (2\pi)^{3/2} W^{(1)} 
 \zeta(\bm{k}=0)\,,
$$
can affect only on the $\bm{k}=0$ mode. Here, $\zeta(\bm{k})$ denotes
the Fourier mode of $\zeta(\bm{x})$. As long as we consider the
 tree-level diagrams, we can neglect the contribution of $W^{(1)}$,
 because we can only observe the correlators of $\zeta(\bm{k})$ with
 $\bm{k} \neq 0$, which will not be affected by $W^{(1)}$ at this order.

\subsection{The power spectrum}
The power spectrum of $\zeta$ is given by using the inverse matrix of
$W^{(2)}(\bm{x}_1,\, \bm{x}_2)$ as
\begin{align}
 & \langle \zeta(\bm{x}_1) \zeta(\bm{x}_2) \rangle_{\rm conn}  =
 \WI (\bm{x}_1,\, \bm{x}_2)\,, \label{Exp:2pzeta}
\end{align}
where Eqs.~(\ref{Def:Wn}) and (\ref{Exp:WQF}) give
\begin{align}
 W^{(2)}(\bm{x}_1,\, \bm{x}_2) = - 2\, {\rm Re} \left[  (\lambda u)^2
 \langle O(\bm{x}_1)  O(\bm{x}_2) \rangle_u \right]  \,. \label{Exp:W2O0} 
\end{align}
Assuming the global translation invariance and the rotational invariance, we express
$W^{(2)}(\bm{x}_1,\, \bm{x}_2)$ in the Fourier space as
\begin{align}
  & W^{(2)}(\bm{x}_1,\, \bm{x}_2) = \int \frac{\dd^3 \bm{k}}{(2\pi)^3}\, e^{i
 \sbm{k} \cdot (\sbm{x}_1- \sbm{x}_2)}\, \hat{W}^{(2)} (k)\,,
\end{align}
where $k\equiv |\bm{k}|$. Using the Fourier mode $\hat{W}^{(2)}(k)$, the power spectrum of the
curvature perturbation is given by
\begin{align}
 & \langle \zeta (\bm{k}_1) \zeta(\bm{k}_2) \rangle_{\rm conn} =
 (2\pi)^3 \delta (\bm{k}_1 + \bm{k}_2) P(k_1) \label{Exp:W2kF}
\end{align}
with 
\begin{align}
 & P(k) = \frac{1}{\hat{W}^{(2)}(k)}\,.
\end{align}

In the deformed conformal field theory, we can yield the almost scale
invariant spectrum which is consistent with the observed spectrum of the
cosmic microwave background. Here, for an illustrative purpose, we
simply approximate the two-point function of $O$ in the right hand
side of Eq.~(\ref{Exp:W2O0}) by the two-point function for the
CFT on the three dimensional Euclidean space $\mathbb{R}^3$ as
\begin{align}
 & W^{(2)}(\bm{x}_1,\, \bm{x}_2) \simeq -2  \, {\rm Re} \left[ (\lambda
 u)^2 \langle O(\bm{x}_1) O(\bm{x}_2) \rangle_{u = 0}
 \right] \,. \label{Exp:W2O} 
\end{align}
In the conformal field theory on $\mathbb{R}^3$ , the Ward-Takahashi
identity for the Weyl transformation determines the two-point function of
$O(\bm{x})$ as
\begin{align}
 & \langle O(\bm{x}_1) O(\bm{x}_2) \rangle_{u=0} = 
 \frac{c}{|\bm{x}_1-\bm{x}_2|^{2\Delta}}\,,
\end{align}
leaving a constant parameter $c$.  (For a review of the conformal field theory on
$\mathbb{R}^3$, see for example, Refs.~\cite{CFT, Emil, YN}.)
In the limit $u \to 0$, the QFT operator $O$ becomes the marginal
operator with $\Delta=3$, because $\langle O O \rangle$ is proportional to 
$\langle T T \rangle$ and the energy momentum tensor becomes a marginal
operator in this limit. We can interpret the constant parameter
$c$ as the central charge, which is a measure of the number of degrees of
freedom that contribute to the marginal deformation. 
Performing the Fourier transformation, we obtain the Fourier mode
$\hat{W}^{(2)} (k)$ as
\begin{align}
 & \hat{W}^{(2)} (k) = - \frac{\pi^2}{6} (\lambda u)^2 c  k^3 \,. \label{Exp:W2k}
\end{align}

Now, we can obtain the power spectrum of the curvature perturbation
$\zeta$ in the cosmological spacetime. Using Eqs.~(\ref{Exp:2pzeta}) and
(\ref{Exp:W2kF}), we can obtain the power spectrum as
\begin{align}
 & P(k)  = - \frac{6}{\pi^2}
 \frac{1}{(\lambda u)^2 c} \frac{1}{k^3}
\,. \label{P} 
\end{align}
We should emphasize that
Eq.~(\ref{P}) is valid irrespective of the strength of the gravitational
coupling in the bulk. In the context of dS/CFT, Strominger determined the
central charge by computing  the trace of the boundary energy momentum tensor
~\cite{Strominger, Strominger2} with the result
\begin{align}
 & c \sim - (\Mp R_{\rm dS})^2, \label{Exp:c}
\end{align}
where $R_{\rm dS}$ is the de Sitter radius. The minus sign relative to the result in Ref. \cite{Strominger}
stems from the fact that there the energy-momentum tensor is defined by taking the
derivative of the bulk action $S_{\rm bulk}$ with respect to the boundary metric. This differs from the notation
we are using here by a factor of $i$, since in the semiclassical limit
$\WQ \sim -i S_{\rm bulk}$ (see Eq.~(\ref{Exp:Ss})).

When we assume the Friedmann equation as the bulk evolution equation and neglect the
quantum corrections to the beta function, the
beta function $\beta$ is given in terms of the slow-roll parameter $\varepsilon$ as
\begin{align}
 & \beta^2 \simeq (\lambda u)^2 \simeq 2 \varepsilon \label{Exp:beta}
\end{align}
at the leading order of the deformation from the conformal field theory
~\cite{vdS, Shiu}. If we use these
expressions (\ref{Exp:c}) and (\ref{Exp:beta}),  Eq.~(\ref{P})
reproduces the well-known power
spectrum obtained in a weakly-coupled inflation driven by a single scale 
field:
\begin{align}
 & P(k) \propto \frac{1}{\varepsilon} \left( H
 \over \Mp \right)^2 \frac{1}{k^3}
\end{align}
with $H \approx  1/R_{\rm dS}$.

\subsection{The bi-spectrum}
\begin{figure}[t]
\begin{center}
\begin{tabular}{cc}
\includegraphics[width=7cm]{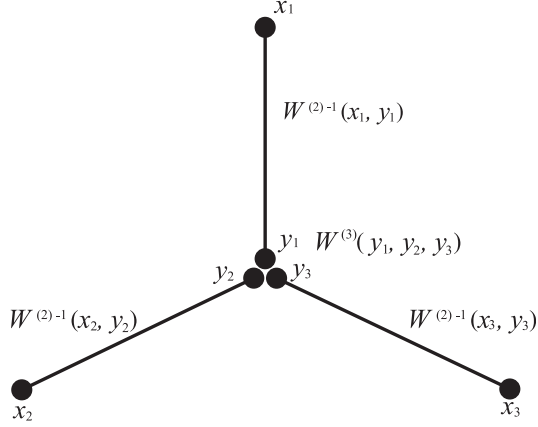}
\end{tabular}
\caption{The diagram of the bi-spectrum}  
\label{Fg:V3}
\end{center}
\end{figure} 
Next, we calculate the non-Gaussian spectrums of the primordial
curvature perturbation $\zeta(\bm{x})$. The bi-spectrum for $\zeta(\bm{x})$ is expressed
by the cubic interaction $W^{(3)}(\bm{x}_1,\, \bm{x}_2,\, \bm{x}_3)$ as
\begin{align}
 & \langle \zeta(\bm{x}_1) \zeta(\bm{x}_2) \zeta(\bm{x}_3) \rangle_{\rm conn} = - \int
 \prod_{i=1}^3  \dd^3 \bm{y}_i\, \WI(\bm{x}_i\,,\bm{y}_i)\,
 W^{(3)} (\bm{y}_1,\,\bm{y}_2,\,\bm{y}_3)\,,   \label{Exp:BSzeta}
\end{align} 
where using Eqs.~(\ref{Def:Wn}) and (\ref{Exp:WQF}), we obtain
\begin{align}
 &W^{(3)}(\bm{x}_1,\, \bm{x}_2,\, \bm{x}_3) = - 2 \, {\rm Re} \Bigl[ (\lambda u)^3 \langle O(\bm{x}_1)
 O(\bm{x}_2) O(\bm{x}_3) \rangle_u  \cr
 & \qquad \qquad  \qquad \qquad  \qquad \qquad  - \lambda^3 u^2 \{\delta (\bm{x}_1 - \bm{x}_2) \langle 
 O(\bm{x}_2) O(\bm{x}_3) \rangle_u + ({\rm 2\, \cp}) \} \Bigr]\,. \cr
\label{Exp:W3} 
\end{align}
In Eq.~(\ref{Exp:BSzeta}), we noted that 
$W^{(3)}(\bm{x}_1,\, \bm{x}_2,\,\bm{x}_3)$ is symmetric under an exchange
of the arguments $\bm{x}_1$, $\bm{x}_2$, and $\bm{x}_3$. The expression
of Eq.~(\ref{Exp:BSzeta}) can be diagrammatically understood as in
Fig.~\ref{Fg:V3}. Performing the Fourier transformation, the bi-spectrum for
$\zeta(\bm{k})$ is given by 
\begin{align}
  \langle \zeta(\bm{k}_1) \zeta(\bm{k}_2) \zeta(\bm{k}_3)
 \rangle_{\rm conn}
 &= (2\pi)^3 \delta (\bm{k}_1 + \bm{k}_2 + \bm{k}_3)\,
 B\left(\bm{k}_1,\, \bm{k}_2,\, \bm{k}_3 \right) \,  \label{Exp:bispzeta}
\end{align}
with
\begin{align}
 & B\left(\bm{k}_1,\, \bm{k}_2,\, \bm{k}_3 \right) = -  \frac{ \hat{W}^{(3)} \left(\bm{k}_1,\, \bm{k}_2,\,
 \bm{k}_3 \right)}{\hat{W}^{(2)}(k_1) \hat{W}^{(2)}(k_2)
 \hat{W}^{(2)}(k_3)} = - \hat{W}^{(3)} \left(\bm{k}_1,\, \bm{k}_2,\,
 \bm{k}_3 \right) \prod_{i=1}^3 P(k_i)\,,  \label{Exp:bispzeta2}
\end{align}
where $\hat{W}^{(3)} \left(\bm{k}_1,\, \bm{k}_2,\, \bm{k}_3 \right)$ is
defined as
\begin{align}
 &(2\pi)^3 \delta (\bm{k}_1 + \bm{k}_2 + \bm{k}_3)
 \hat{W}^{(3)} \left(\bm{k}_1,\, \bm{k}_2,\, \bm{k}_3 \right)  \equiv
   \prod_{i=1}^3 \int \dd^3 \bm{x}_i\, e^{-i \sbm{k}_i
 \cdot \sbm{x}_i} W^{(3)}(\bm{x}_1,\, \bm{x}_2,\,\bm{x}_3)\,.
\end{align}
The bi-spectrum (\ref{Exp:bispzeta}) agrees with the one obtained in Ref.~\cite{BMS} by using the
holographic renormalization group method.

\subsection{The tri-spectrum}
\begin{figure}[t]
\begin{center}
\begin{tabular}{cc}
\includegraphics[width=14cm]{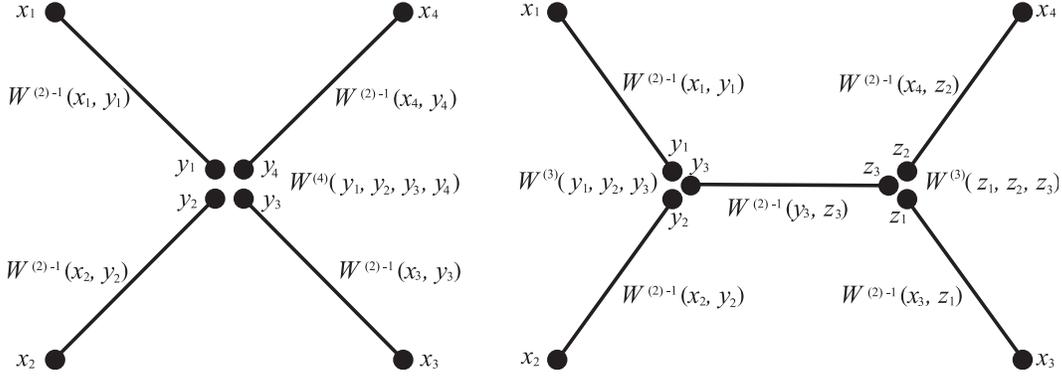}
\end{tabular}
\caption{The diagrams of the tri-spectrum}  
\label{Fg:V4}
\end{center}
\end{figure}

Next, we calculate the tri-spectrum of $\zeta(\bm{x})$. As is depicted in
Fig.~\ref{Fg:V4}, the tri-spectrum is composed of the two-different diagrams and is
given by
\begin{align}
  & \langle \zeta(\bm{x}_1) \zeta(\bm{x}_2) \zeta(\bm{x}_3)
 \zeta(\bm{x}_4) \rangle_{\rm conn} \cr
  &= - \int
  \prod_{i=1}^4 \dd^3 \bm{y}_i\, \WI(\bm{x}_i\,,\bm{y}_i)\,
 W^{(4)} (\bm{y}_1,\,\bm{y}_2,\,\bm{y}_3,\,\bm{y}_4) \cr
 & \qquad   + \int \prod_{i=1}^3 
 {\rm d}^3 \bm{y}_i \int {\rm d}^3  \bm{z}_i\,\WI(\bm{x}_1,\,\bm{y}_1)
 \WI (\bm{x}_2,\,\bm{y}_2) \cr
 & \qquad \qquad  \times \WI(\bm{x}_3,\,\bm{z}_1) \WI(\bm{x}_4,\,\bm{z}_2)
 \WI (\bm{y}_3,\,\bm{z}_3)  W^{(3)}( \bm{y}_1,\,
  \bm{y}_2,\, \bm{y}_3) W^{(3)}(\bm{z}_1,\,\bm{z}_2,\,\bm{z}_3)  \cr
 & \qquad + ( \bm{x}_2 \leftrightarrow \bm{x}_3) + ( \bm{x}_2 \leftrightarrow \bm{x}_4)\,,
\end{align}
where the first term stems from the diagram in the left panel of
Fig.~\ref{Fg:V4} and the second term stems from the diagram in the
right. 
The terms in the last lines
are the contributions from the diagram with the vertex $\bm{x}_2$ in
the right panel of Fig.~\ref{Fg:V4} exchanged with $\bm{x}_3$ and $\bm{x}_4$,
respectively. Here, noticing the fact that $W^{(3)}(\bm{x}_1,\, \bm{x}_2,\, \bm{x}_3)$
and $W^{(4)}(\bm{x}_1,\, \bm{x}_2,\, \bm{x}_3,\, \bm{x}_4)$ are
symmetric under an exchange of their arguments, we summed up the
terms which give the same contribution. Equations ~(\ref{Def:Wn}) and (\ref{Exp:WQF}) yield  
\begin{align}
 & W^{(4)} (\bm{x}_1,\,\bm{x}_2,\,\bm{x}_3,\,\bm{x}_4) \cr
 & = - 2 \, {\rm Re} \biggl[ (\lambda u)^4 \langle O(\bm{x}_1) \cdots 
 O(\bm{x}_4) \rangle_u \hspace{-2pt} - \lambda^4 u^3\! \left\{ \delta
 (\bm{x}_1 - \bm{x}_2)  \langle O(\bm{x}_2) O(\bm{x}_3) 
 O(\bm{x}_4) \rangle_u  + \left( 5\,\,\cp \right) \! \right\} \cr
 & \qquad  \qquad \quad    + \lambda^4 u^2 \{  \delta (\bm{x}_1 - \bm{x}_2) \delta(\bm{x}_2 -
 \bm{x}_3) \langle O(\bm{x}_3) O(\bm{x}_4) \rangle_u +
 (3\,\, {\rm \cp}) \} \biggr]\,.
\end{align}
In the Fourier space, the tri-spectrum is given by
\begin{align}
 \langle \zeta(\bm{k}_1) \zeta(\bm{k}_2) \zeta(\bm{k}_3) \zeta(\bm{k}_4)
 \rangle_{\rm conn} &= (2\pi)^3\, \delta\left( \sum_{i=1}^4 \bm{k}_i\right) \,
  T \left( \bm{k}_1,\,\bm{k}_2,\, \bm{k}_3,\, \bm{k}_4 \right)\,,
\end{align}
with
\begin{align}
 & T (\bm{k}_1,\, \bm{k}_2,\, \bm{k}_3,\, \bm{k}_4) = T_1 \left( \bm{k}_1,\,\bm{k}_2,\, \bm{k}_3,\, \bm{k}_4 \right) +
 T_2 \left( \bm{k}_1,\,\bm{k}_2,\, \bm{k}_3,\, \bm{k}_4 \right) \cr
 & \qquad \qquad \qquad \qquad \qquad \qquad   + T_2
 \left( \bm{k}_1,\,\bm{k}_3,\, \bm{k}_2,\, \bm{k}_4 \right) + T_2 \left(
 \bm{k}_1,\,\bm{k}_4,\, \bm{k}_3,\, \bm{k}_1 \right) \,,  \label{Def:T} \\
 & T_1(\bm{k}_1,\, \bm{k}_2,\, \bm{k}_3,\, \bm{k}_4)  = - 
 \hat{W}^{(4)} \left(\bm{k}_1,\, \bm{k}_2,\,\bm{k}_3,\, \bm{k}_4 \right)
  \prod_{i=1}^4 P(k_i) \,, \\
  & T_2(\bm{k}_1,\, \bm{k}_2,\, \bm{k}_3,\, \bm{k}_4)  = \hat{W}^{(3)}\left(
 \bm{k}_1,\, \bm{k}_2,\, - \bm{k}_{12} \right) \hat{W}^{(3)}\left(
 \bm{k}_3,\, \bm{k}_4,\, - \bm{k}_{34} \right) P (k_{12}) \prod_{i=1}^4 P(k_i)   \,, \label{Exp:T} 
\end{align}
where we introduced the Fourier mode of $W^{(4)}$ as
\begin{align}
  (2\pi)^3 \hat{W}^{(4)} \left(\bm{k}_1,\, \bm{k}_2,\, \bm{k}_3,\, \bm{k}_4
 \right)  \delta\left( \sum_{i=1}^4 \bm{k}_i\right)  \equiv
   \prod_{i=1}^4 \int \dd^3 \bm{x}_i\, e^{-i \sbm{k}_i
 \cdot \sbm{x}_i} W^{(4)}(\bm{x}_1,\, \bm{x}_2,\,\bm{x}_3,\, \bm{x}_4)\,,
\end{align}
and the momentum $\bm{k}_{ij}$ and its absolute value as
$\bm{k}_{ij}\equiv \bm{k}_i + \bm{k}_j$ and $k_{ij}\equiv |\bm{k}_{ij}|$.
Note that using the bi-spectrum $B(\bm{k}_1,\, \bm{k}_2,\, \bm{k}_3)$,
we can express $ T_2(\bm{k}_1,\, \bm{k}_2,\, \bm{k}_3,\, \bm{k}_4)$ as
\begin{align}
 &  T_2(\bm{k}_1,\, \bm{k}_2,\, \bm{k}_3,\, \bm{k}_4)  =  \frac{B \left(
 \bm{k}_1,\, \bm{k}_2,\, - \bm{k}_{12} \right)  B \left(
 \bm{k}_3,\, \bm{k}_4,\, - \bm{k}_{34} \right)}{P(k_{12})}\,.
 \label{Exp:T2} 
\end{align}
An extension to higher-point correlators proceeds in a straightforward
manner. Thus, once we obtain all the $m$-point functions for the QFT
operator $O(\bm{x})$ with $m \leq n$, we can easily obtain the
$n$-point functions for the curvature perturbation
$\zeta(\bm{x})$. Note that although the computation of the $n$-point
functions proceeds as in the perturbative expansion of the weakly
coupled QFT, the propagator 
$\WI(\bm{x}_1,\, \bm{x}_2)$ and the vertex functions 
$W^{(n)}(\bm{x}_1,\, \cdots,\, \bm{x}_n)$ with $n\geq 3$ can include the resumed
non-perturbative effect of the boundary QFT.

\subsection{The Suyama-Yamaguchi inequality}
In the conventional inflation, where the gravity is weakly coupled and
the perturbative analysis in the bulk is valid, the
non-linear parameters $f_{\rm NL}$ and $\tau_{\rm NL}$: 
\begin{align}
 & f_{\rm NL} \equiv \frac{5}{12} \lim_{k_1 \to 0}  \frac{B(\bm{k}_1,\, \bm{k}_2,\,
 -(\bm{k}_1 + \bm{k}_2))}{P(k_1) P(k_2)}  \\
 & \tau_{\rm NL} \equiv \frac{1}{4} \lim_{k_{12} \to 0}  \frac{T(\bm{k}_1,\, \bm{k}_2,\,
 \bm{k}_3,\, \bm{k}_4)}{P(k_1) P(k_3) P(k_{12})}
\end{align}
are known to satisfy the so-called Suyama-Yamaguchi (SY) inequality~\cite{SY}:
\begin{align}
 & \tau_{\rm NL} \geq \left( \frac{6}{5} f_{\rm NL} \right)^2\,. \label{SY}
\end{align}
The SY inequality is first shown
by using the $\delta N$ formalism, which gives a map between the field
fluctuation at the Hubble crossing time and the fluctuation of the
number of $e$-folding that approximately agrees with the curvature
perturbation at the end of inflation, under the assumption that the field
fluctuation is totally Gaussian at the Hubble crossing time. Under this
assumption, the equality holds for a single field model of inflation. 
A generalization of the SY inequality has been intended later, for
instance in Refs.~\cite{Riotto, ABG}. The authors of Ref.~\cite{ABG}
showed that if $\tau_{\rm NL}$ and $(f_{\rm NL})^2$ are either momentum
independent or have the same momentum dependence, we can verify the SY
inequality as well as in the presence of the non-Gaussianity at the
Hubble crossing time.

Here, we study the validity of this inequality in holographic
inflation. As the limit $k_{12} \to 0$, more precisely, we mean the
limit $k_{12} \ll k_1, k_3, k_{13}$. Using Eq.~(\ref{Exp:T2}), we obtain the
contribution of the second term in Eq.~(\ref{Def:T}) to $\tau_{\rm NL}$ as 
\begin{align}
 &   \lim_{k_{12} \to 0}  \frac{T_2(\bm{k}_1,\, \bm{k}_2,\,
 \bm{k}_3,\, \bm{k}_4)}{P(k_1) P(k_3) P(k_{12})} \to 4 \left( \frac{6}{5}
 f_{\rm NL} \right)^2 \,.
\end{align}
The third term in Eq.~(\ref{Def:T}) gives
\begin{align}
 &   \frac{T_2(\bm{k}_1,\, \bm{k}_3,\, \bm{k}_2,\, \bm{k}_4)}{P(k_1)
 P(k_3) P(k_{12})} = \frac{P(k_{13})}{P(k_{12})} \left\{
 \frac{B(\bm{k}_1,\,-\bm{k}_{13}, \,\bm{k}_3)}{P(k_1) P(k_{13})}
 \right\}^2 \,, \label{T2tau}
\end{align}
where we noted that
$\bm{k}_1 \simeq - \bm{k}_2$ and $\bm{k}_3 \simeq - \bm{k}_4$ in the limit $k_{12} \to 0$. 
The term in the curly
brackets is expressed by $k_1/k_3$ or $k_3/k_1$ which are larger than
$k_{12}/k_1$ and $k_{12}/k_3$. Therefore the right hand side of
Eq.~(\ref{T2tau}) with $1/P(k_{12})$, which approaches 0
in the limit $k_{12} \to 0$, vanishes. The fourth term in Eq.~(\ref{Def:T}) also vanishes in this
limit. Thus, we obtain
\begin{align}
 & \tau_{\rm NL} - \left( \frac{6}{5} f_{\rm NL} \right)^2 = - \frac{1}{4}
 \lim_{k_{12} \to 0} \hat{W}^{(4)} \left(\bm{k}_1,\, \bm{k}_2,\, \bm{k}_3,\, \bm{k}_4
 \right) \frac{P(k_1) P(k_3)}{P(k_{12})}\,, \label{Eq:SY}
\end{align}
and hence the SY inequality holds, if the right hand side of
Eq.~(\ref{Eq:SY}) is equal to or is larger than $0$. If the vertex
function 
$\hat{W}^{(4)}(\bm{k}_1,\, \bm{k}_2,\, \bm{k}_3,\, \bm{k}_4)$ does not yield a
singular pole in the limit $k_{12} \to 0$, the right hand side of
Eq.~(\ref{Eq:SY}) vanishes and hence the equality holds as in single
field models of the weakly coupled inflation. However, the
momentum dependence of the vertex function 
$\hat{W}^{(4)}$ which includes the
non-perturbative effect, is rather less obvious. To capture 
the momentum dependence of the vertex function 
$\hat{W}^{(4)}$, we need
to specify the four-point function of $O(\bm{x})$.  As is known,
even in CFT, the higher point functions 
$\langle O(\bm{x}_1) \cdots O(\bm{x}_n) \rangle_{u=0}$
with $n\geq 4$ are not determined only by the conformal symmetry. Actually, the Ward-Takahashi identities leave functional degree of
freedom in the four point function~\cite{Emil}. It would be interesting
to examine the validity of the SY inequality in a particular model of the
holographic inflation.

\section{Concluding remarks} \label{Sec:Conclusion}
In this paper, using the holographic formula (\ref{Exp:psi}) and the
Ward-Takahashi identity, we provided the relation between the $n$-point
functions of the bulk gravitational field $\zeta$ and $n$-point functions
of the QFT operator $O$ on the boundary. In contrast with
Refs.~\cite{BMS, Shiu}, we expressed the Ward-Takahashi identity so that it
directly relates the generating functional $Z_{\rm QFT}$ to the $n$-point
functions of the QFT operator $O$. This bypasses a redundant step which refers to the energy-momentum tensor,
making the computation more compact and transparent. This result can be also used to compute the spectra of
$\zeta$ based on the domain wall/cosmology correspondence, which is addressed in
Refs.~\cite{MS_HC09, MS_HC10, MS_HCob10, MS_NG, MS_NGGW, BMS}. In these papers, 
the description of the spectra is provided based on the holographic renormalization
method. We showed that both methods lead to the same formula relating the $n$-point
functions of $\zeta$ and those of $O$.  In Refs.~\cite{MS_HC09,
MS_HC10, MS_NGGW}, the correlators of the tensor modes are also
studied. An extension of our argument to include these correlators is
left for a future study. We also
studied the validity of the Suyama-Yamaguchi inequality and showed that
unless $W^{(4)}$ yields a singular pole in the limit $k_{12} \to 0$,
$\tau_{\rm NL}$ agrees with $(6 f_{\rm NL}/5)^2$ as has been known in
weakly coupled single field models.

When we calculate the $n$-point functions of the curvature perturbation
$\zeta$, using the vertex functions 
$W^{(n)}(\bm{x}_1, \cdots,\,\bm{x}_n)$, we neglected the loop
corrections. (Note that the vertex function 
$W^{(n)}(\bm{x}_1, \cdots,\, \bm{x}_n)$ includes the non-perturbative
effect of the QFT on the boundary.) As a final issue of the present
article, we examine the validity of this approximation. Loop diagrams
can be obtained by inserting additional interaction vertices to the
tree level diagrams, presented in Figs.~\ref{Fg:V3} and \ref{Fg:V4}. 
A naive order estimation tells that inserting a further $n$-point
interaction vertex
$$
\prod_{i=1}^n \int \dd^3 \bm{x}_i \zeta(\bm{x}_i)  W^{(n)}(\bm{x}_1,\,
\cdots ,\, \bm{x}_n)
$$
changes the amplitude by
$$
|W^{(n)} \zeta^n| = {\cal O}(\lambda^n u^m) \times 
{\cal O} \left( (\lambda u c^{1/2})^{-n} \right) = {\cal O} (u^{m-n} c^{-n/2})\,,
$$
with $1\leq m \leq n$. Here, we noted that as is suggested by the order of the propagator given
in Eq.~(\ref{P}), the amplitude of $\zeta$ amounts to 
$1/(\lambda u c^{1/2})$ and the contribution of the
$n$-th vertex function amounts to $\lambda^n u^m$. Therefore, to keep the loop corrections sub dominant, the
central charge should be $c \gg 1$. Since the central charge is supposed to be the number of
degrees of freedom, this condition can be thought of as the
condition of the so-called large $N$ approximation. When we use Eq.~(\ref{Exp:c}), $c \gg 1$ yields
$(H/\Mp) \ll 1$, which requests that inflation should take place in the
sub-Planck energy scale. Note that the energy scale of inflation, $H$,
can exceed the string scale, admitting the stringy corrections.

\acknowledgments
This work is supported by the Grant-in-Aid for the Global COE Program
"The Next Generation of Physics, Spun from Universality and Emergence"
from the Ministry of Education, Culture, Sports, Science and Technology
(MEXT) of Japan. J.~G. and Y.~U. are partially supported by MEC
FPA2010-20807-C02-02, AGAUR 2009-SGR-168 and CPAN CSD2007-00042
Consolider-Ingenio 2010. We thank B.~Fiol and K.~Skenderis for their
valuable comments.


\end{document}